\documentclass[conference]{IEEEtran}
\IEEEoverridecommandlockouts
\usepackage{cite}
\usepackage{graphicx}
\usepackage{amsmath,amsfonts,amssymb,amsthm}
\usepackage[dvipsnames]{xcolor}
\usepackage{array,multirow}
\usepackage[english]{babel}
\usepackage{textcomp}
\usepackage{url}
\usepackage[linesnumbered,ruled,vlined]{algorithm2e}

\usepackage[caption=false,font=footnotesize]{subfig}
\begin{document}
\title{Critical Weather Scenario Screening Using Weather-to-Voltage (W2V) Predictive Modeling\thanks{This work has been supported by NSF grants 2130706 and 2150571, and by UT Austin-Amazon Science Hub.}}

\author{
\IEEEauthorblockN{Sol Lim, Min-Seung Ko, and Hao Zhu}
\IEEEauthorblockA{
Chandra Family Department of Electrical and Computer Engineering\\
The University of Texas at Austin, Austin, TX, USA\\
\{sollim, kms4634500, haozhu\}@utexas.edu
}
}

\maketitle

\begin{abstract}
This paper proposes a critical weather scenario screening framework for identifying weather conditions that can trigger 
high-voltage (HV) events in the power grid. Unlike conventional weather-aware contingency analysis limited to component-level outage risk, our framework screens weather scenarios as potential drivers of grid-level voltage violations. Given a non-critical weather scenario,
we seek the perturbation over the high-dimensional weather space to maximize a pre-defined voltage criticality score, by using a differentiable weather-to-voltage (W2V) predictive model to facilitate the gradient update over a compact latent space. 
Specifically, a non-negativity constraint is used for achieving physically-consistent perturbations, with another L1-norm based constraint for bounded perturbation. The latter could promote sparse and interpretable perturbations, and this uniform budget also yields a sensitivity-aware vulnerability ranking across different weather scenarios. 
Numerical experiments on a 6717-bus synthetic Texas system have effectively demonstrated the potential of weather uncertainty in triggering HV events, and this potential cannot be represented by the voltage analysis of individual weather scenarios. Interestingly, the most vulnerable scenarios are characterized by wind-dominated perturbation patterns concentrated in high wind-capacity regions, coinciding with observations from actual power flow data and experiences in real system operations.
\end{abstract}


\section{Introduction}
\label{sec:introduction}
The rapid expansion of weather-based renewable generation has strengthened the coupling between meteorological conditions and power system operating states. Weather conditions such as wind speed, solar irradiance, and temperature exhibit a strong spatial correlation and their high uncertainty can jointly affect renewable generation and load demand. As these correlated changes propagate through grid power flows, they can collectively worsen system-wide voltage deviations~\cite{bloom2016eastern, vittal2009steady}. With increasing renewable penetration, identifying the grid operational risks induced by weather variability has become an important requirement for reliable power delivery~\cite{panteli2015influence}.

Current studies on weather-induced grid risk have primarily focused on contingency analysis. For example, weather-dependent renewable variability has been incorporated into the $N-1$, or even $N-k$, contingency assessment to evaluate the impact of different generation failure scenarios~\cite{jiang2016power, stover2025probabilistic}. In recent regulatory changes, FERC Order 896~\cite{ferc896} mandates the consideration of extreme weather events in grid reliability assessments. Meanwhile, EPRI has proposed a probabilistic framework to integrate the uncertainty of renewable generation into the statistical analysis of equipment outages~\cite{epri2025risk}, among other industry efforts. These recent approaches could support real-world grid operations, but a common issue remains: when incorporating weather as an exogenous modifier of contingency severity or likelihood of failure, such direct effect is still limited to component-level outages. 

However, the large spatial correlation of weather events could have a wide-area impact on grid operating conditions even in the absence of component outages, as mentioned earlier. This inspires a new screening problem to identify weather scenarios under which weather uncertainty could induce grid reliability issues such as voltage violations. To the best of our knowledge, a systematic framework for such weather scenario screening remains insufficiently explored.
Addressing this gap requires a model that can efficiently evaluate how various scenarios of weather uncertainty propagate to grid-wide voltage responses. Existing weather-incorporated power flow approaches~\cite{overbye2023approach,safdarian2024calculation}, while accurate, rely on simulators and therefore cannot directly support analytical formulation and optimal search. The weather-to-voltage (W2V) predictive model developed in~\cite{nexus_paper} addresses this limitation by providing a differentiable neural surrogate that maps spatial weather features to grid-wide voltages. In this work, we further strengthen the W2V model design in its loss function and initialization scheme, particularly improving the prediction accuracy in the high-voltage (HV) regime that is central to the screening task.

Building on the improved W2V model, we formulate weather scenario screening as a constrained perturbation optimization problem in the compact W2V latent space. Given a non-critical weather scenario, we search for a standardized perturbation that maximizes a pre-defined voltage criticality score. An L1-norm budget bounds the overall perturbation magnitude to promote a small number of dominant latent directions for sparse and interpretable solutions.
The resulting latent perturbation is mapped back to the weather space, where non-negativity constraints maintain the physical consistency of the perturbed weather. By applying the same perturbation budget across candidate scenarios, the optimized criticality scores can be compared on a consistent basis. Accordingly, a scenario with a larger optimized score is regarded as more vulnerable to HV violations since its local weather neighborhood contains conditions that induce greater voltage stress. In this way, the proposed screening goes beyond evaluating the initial criticality of weather scenarios and identifies HV vulnerabilities that become apparent only under targeted weather perturbations.
Our main contributions are summarized as follows:
\begin{enumerate}
    \item A critical weather scenario screening framework is proposed to identify weather conditions that are prone to HV violations under bounded weather uncertainty based on a differentiable W2V predictive model.
 
    \item Our constrained optimization problem seeks to identify perturbations to any given operating point under an L1-norm budget, and a non-negativity constraint to ensure physically-consistent solutions, efficiently solvable via a primal-dual method with L1 projection.

    \item We validate the proposed framework on a 6717-bus synthetic Texas grid with realistic weather data, showing the potential of weather uncertainty to trigger HV events beyond what the initial criticality scores suggest. The most vulnerable scenarios exhibit wind-dominated perturbations concentrated in regions with high wind generation capacity, consistent with the HV events from the original power flow dataset.
\end{enumerate}

\section{W2V-Based Weather Scenario Screening}
\label{sec:prob_formulation}
This section presents the formulation of the W2V-based weather scenario screening problem. We first summarize the development of the W2V predictive model with a focus on improving high-voltage prediction, and then formulate the constrained perturbation problem.

\subsection{Weather-to-Voltage (W2V) Predictive Model}
\label{sec:w2v_model}
The W2V model is an autoencoder-based neural surrogate that maps spatial weather features in $\mathbf{w} \in \mathbb{R}^W$ to grid-wide bus voltages in $\mathbf{v} \in \mathbb{R}^B$ through a low-dimensional latent representation $\mathbf{z} \in \mathbb{R}^K$. The weather input vector consists of temperature, wind speed, and global horizontal irradiance (GHI) measured across geographically distributed locations. The voltage vector represents the quasi-steady state power flow response corresponding to the weather conditions, obtained by measurements or simulators~\cite{overbye2023approach}. Thus, both input and output capture the snapshot view of weather and voltage, without any temporal dynamics.

In the W2V model, an encoder $E$ maps the input to a latent vector $\mathbf{z}$, followed by the voltage decoder $D_v$ to form the predicted voltage $\hat{\mathbf{v}}$,  given by:
\begin{align}
\mathbf{z} = E(\mathbf{w}), \quad \hat{\mathbf{v}} = D_v(\mathbf{z}).
\label{eq:w2v_map}
\end{align}
The encoder and voltage decoder are jointly trained with an auxiliary weather reconstruction decoder $D_w$ that regularizes the latent space to preserve meteorological information. Accordingly, the training objective combines both errors as:
\begin{align}
\mathcal{L} = \mathcal{L}_v(\hat{\mathbf{v}}, \mathbf{v}) + \lambda\,\mathcal{L}_w(\hat{\mathbf{w}}, \mathbf{w}),
\label{eq:w2v_loss}
\end{align}
where the loss functions $\mathcal{L}_v$ and $\mathcal{L}_w$ are balanced by the hyperparameter $\lambda > 0$. Full architectural and training details are provided in~\cite{nexus_paper}. Although the standard mean-square error (MSE) is a fine choice for loss functions, it may not be effective for predicting the HV samples which are the focus of this paper. This is because MSE tends to emphasize the dominant samples at the nominal-voltage region and the occurrence of HV events is rare~\cite{yang2021delving,rudy2023output}. To improve the prediction accuracy in the HV tail, this work adopts a new voltage-weighted loss, given by:
\begin{align}
\mathcal{L}_v = \tfrac{1}{B}\sum_{i=1}^{B} \phi(v_{i})\,(\hat{v}_{i} - v_{i})^2,
\label{eq:mmse}
\end{align}
where subscript $i$ denotes the bus index. The weight function 
$\phi(v) = 1 + \alpha \bigl(\tfrac{v - v_{\min}}{v_{\max} - v_{\min}}\bigr)^{p}$ smoothly ramps from $1$ to $(1 + \alpha)$ within the range of $[v_{\min},v_{\max}]$. Hyperparameters $\alpha > 0$ and $p$ manage the strength and sharpness of the HV emphasis. This new modification of the voltage loss makes the resultant W2V model better match the HV events of interest in this work. 

As our screening task will directly perturb the latent $\mathbf{z}$, we discuss the design of the encoder in more detail here.  
For simplicity, we use an affine layer for the encoder, which makes the mapping from $\mathbf{z}$ to weather $\mathbf{w}$ very convenient. The encoder is initialized by the principal component analysis (PCA) on the full input data matrix to provide a numerically stable low-dimensional representation. 
To better capture meteorological correlations, we further combine feature-specific principal components (PCs) with additional shared PCs obtained from a joint PCA of temperature and GHI, which are strongly coupled through diurnal solar radiation patterns. Wind speed, which exhibits a different spatial structure, retains its own feature-specific PCs. The resulting $K$ PCs in total provide a richer warm-start that reflects both feature-specific variability and cross-feature correlation. 

We want to highlight two properties of our W2V model that can facilitate the ensuing screening task. First, the decoder $D_v$ is differentiable with respect to $\mathbf{z}$, enabling gradient-based search for voltage-stressing perturbations, as discussed soon. Second, the compact latent representation provides a lower-dimensional search space for perturbation than the original weather input, benefiting the screening problem with high computational tractability. 

Beyond these advantages, the W2V model design also offers flexibility for practical deployment. Although the model is trained using simulation-generated data that does not reflect measurement uncertainty in real systems, it can be calibrated or retrained using field measurements. Furthermore, the W2V model does not require an explicit network representation to learn the direct weather-to-voltage mapping. Thus, although the current training data is generated under a fixed network topology, the model can be retrained using weather--voltage data corresponding to a different network topology without modifying its architecture. In either case, once the model is trained, it readily provides the weather-to-voltage sensitivity that underpins the proposed screening task.

\subsection{Constrained Perturbation Optimization}
\label{sec:optimization}

The screening process seeks to identify weather scenarios that can become HV-critical under small but physically admissible perturbations. 
To quantify voltage stress, we first define the \emph{criticality score} as the average of the top-$k$ predicted bus voltages based on the decoder in \eqref{eq:w2v_map}:
\begin{align}
S(\mathbf{z}) = \tfrac{1}{k} \sum_{i \in \mathcal{T}(\mathbf{z})} \hat{v}_i(\mathbf{z}),
\label{eq:S}
\end{align}
where $\mathcal{T}(\mathbf{z})$ is the index set of the $k$ buses with the highest predicted voltages under latent state $\mathbf{z}$.
This score is used instead of the maximum voltage at a single bus because the top-$k$ average provides a smoother and more stable gradient for optimization while still capturing the upper tail of the voltage distribution that characterizes HV events.

Each weather scenario $\mathbf{w}$ is then classified as non-critical if its criticality score satisfies $S(\mathbf{z}) < \gamma$. The screening threshold $\gamma$ is calibrated from the full dataset by labeling a timestep as HV-critical if at least one bus voltage exceeds the voltage threshold $\tau$. The value of $\gamma$ is then chosen to maximize Youden's $J$ statistic~\cite{youden1950index}, which is evaluated along the receiver operating characteristic (ROC) 
curve of $S$ scores, so that the criticality score best separates HV and non-HV samples. Non-critical scenarios near this threshold are used as perturbation candidates since they are close to the HV boundary but do not violate the limit.

Given a non-critical weather scenario $\mathbf{w}$, 
the proposed framework seeks the perturbation that maximizes voltage stress under a fixed perturbation budget. By using the same budget, the attained stress level measures how vulnerable each scenario is to HV triggering. 
Rather than perturbing the high-dimensional weather input directly, we operate in the latent space by perturbing $\mathbf{z}$. Let $\boldsymbol{\sigma}=(\sigma_1,\ldots,\sigma_K)$ denote the vector of empirical standard deviations of the latent variables over the training set. We define the standardized perturbation $\mathbf{u}\in\mathbb{R}^K$ so that the perturbed latent vector is $(\mathbf{z}+\mathrm{diag}(\boldsymbol{\sigma})\,\mathbf{u})$, ensuring that perturbations across latent variables are measured relative to their empirical variability. This weighted perturbation can prevent the budget from being dominated by high-variance dimensions. Since the encoder is affine, the latent and weather perturbations satisfy the linear relation $\mathrm{diag}(\boldsymbol{\sigma})\,\mathbf{u} = \mathbf{A}_E\,\boldsymbol{\delta}_\mathbf{w}$, where $\mathbf{A}_E$ is the encoder weight matrix. 
Accordingly, the corresponding weather perturbation can be recovered exactly from the latent perturbation by
\begin{align}
\boldsymbol{\delta}_\mathbf{w} = \mathbf{A}_E^{\dagger}\,
\mathrm{diag}(\boldsymbol{\sigma})\,\mathbf{u},
\label{eq:delta_w}
\end{align}
where $\mathbf{A}_E^{\dagger}$ denotes the Moore--Penrose pseudo-inverse of $\mathbf{A}_E$.

The constrained perturbation problem is then cast as
\begin{align}
\max_{\mathbf{u}} \quad 
  & S\bigl(\mathbf{z} + \mathrm{diag}(\boldsymbol{\sigma})\,
    \mathbf{u}\bigr) \label{eq:opt_obj} \\
\text{s.t.} \quad 
  & \|\mathbf{u}\|_1 \;\leq\; \epsilon, \label{eq:opt_l1} \\
  & \mathbf{w}_{\geq 0} + \bigl(\boldsymbol{\delta}_{\mathbf{w}}\bigr)_{\geq 0} \;\geq\; \mathbf{0} 
    \label{eq:opt_phys}
\end{align}
where $\mathbf{w}_{\geq 0}$ denotes the sub-vector of non-negative weather features (wind speed and GHI), and $(\cdot)_{\geq 0}$ selects the corresponding entries.
The L1 constraint~\eqref{eq:opt_l1} promotes sparse activation of a few latent variables, yielding interpretable perturbation directions. The non-negativity constraint~\eqref{eq:opt_phys} prevents wind speed and GHI from becoming negative after perturbation. Since the optimizer maximizes the voltage stress without knowing the physical meaning of individual weather features, it can push wind speed and GHI below zero regardless of the perturbation budget $\epsilon$. The constraint is therefore essential for excluding such physically impossible values. Temperature is left unconstrained as it admits negative values. For each candidate scenario, the optimal value $S^\star = S(\mathbf{z} + \mathrm{diag}(\boldsymbol{\sigma})\,\mathbf{u}^\star)$ is used as the vulnerability score. A larger $S^\star$ indicates that the scenario is more vulnerable to HV triggering under the same perturbation budget. The associated recovered weather perturbation $\boldsymbol{\delta}_{\mathbf{w}}^{\star} = \mathbf{A}_E^{\dagger}\,\mathrm{diag}(\boldsymbol{\sigma})\,\mathbf{u}^{\star}$ is used to interpret the weather features and geographical regions that drive the voltage stress.

\section{Primal--Dual Solution with L1 Projection}
\label{sec:solution}
The constrained perturbation problem is non-convex since the voltage decoder is a non-linear neural network and the top-$k$ criticality score depends on the ordering of predicted bus voltages. Nevertheless, the problem is differentiable almost everywhere and can be efficiently solved through gradient-based optimization. The proposed solution combines projected primal updates for the L1 budget with dual updates for the non-negativity constraints.

Since the original formulation maximizes the criticality score, we consider the following partial Lagrangian:
\begin{align}
\mathcal{L}(\mathbf{u}, \boldsymbol{\mu}) 
=& -S\bigl(\mathbf{z} + \mathrm{diag}(\boldsymbol{\sigma})\,
  \mathbf{u}\bigr) 
  + \boldsymbol{\mu}^{\top}\mathbf{g}(\mathbf{u}), \nonumber \\
  &\textrm{with}~\mathbf{g}(\mathbf{u}) = -\mathbf{w}_{\geq 0} - \bigl(\boldsymbol{\delta}_{\mathbf{w}}\bigr)_{\geq 0}
\label{eq:lagrangian}
\end{align}
where $\boldsymbol{\mu} \geq \mathbf{0}$ is the nonnegative dual variable for the constraint in \eqref{eq:opt_phys}. 
 A positive entry of $\mathbf{g}$ indicates the violation of the corresponding non-negativity constraint. The L1 budget is considered an implicit constraint and is therefore not included in~\eqref{eq:lagrangian}. 

\begin{algorithm}[t]
\caption{Primal--Dual Solution with L1 Projection}
\label{alg:pd_l1}
\KwIn{$\mathbf{z}$, $\boldsymbol{\sigma}$, $\epsilon$, step sizes $\eta_{\mathrm{p}}, \eta_{\mathrm{d}}$}
$\mathbf{u} \leftarrow \mathbf{0}$, $\boldsymbol{\mu} \leftarrow \mathbf{0}$, $\mathbf{u}^{\star} \leftarrow \mathbf{u}$\;
\For{$n = 1, 2, \ldots$}{
  $\mathbf{u} \leftarrow \mathrm{Adam}\bigl(\mathbf{u},\, \nabla_{\mathbf{u}} \mathcal{L}\bigr)$ \tcp*{primal gradient descent}
  $\mathbf{u} \leftarrow \Pi_{\|\cdot\|_1 \leq \epsilon}(\mathbf{u})$ \tcp*{L1-ball projection}
  $\boldsymbol{\mu} \leftarrow [\boldsymbol{\mu} + \eta_{\mathrm{d}}\,\mathbf{g}(\mathbf{u})]_{+}$ \tcp*{dual ascent}
  Update $\mathbf{u}^{\star}$ if feasible and $S$ improves\;
  Decay $\eta_{\mathrm{p}}, \eta_{\mathrm{d}}$\;
  \If{no improvement for patience window}{
    \textbf{break}\;
  }
}
\KwOut{$\mathbf{u}^{\star}$}
\end{algorithm}

The resulting algorithm performs three steps per iteration, as summarized in Algorithm~\ref{alg:pd_l1}. The primal variable $\mathbf{u}$ is updated by an Adam gradient step on $\mathcal{L}$, followed by projection onto the L1 ball~\cite{duchi2008efficient}. The dual variable $\boldsymbol{\mu}$ is then updated by projected gradient ascent, where non-negativity is enforced via the $[\cdot]_+$ operator. We set the dual step size larger than the primal step size to ensure rapid enforcement of feasibility. Both step sizes are decayed over iterations to balance early exploration with late-stage refinement. Since this method does not guarantee the global optimality of the non-convex problem, we track the best feasible iterate $\mathbf{u}^\star$---the feasible iterate with the highest $S$---and return it as the final solution.

\section{Experimental Results}
\label{sec:experiment}
This section presents the experimental results of the proposed screening framework. We first describe the experimental setup and compare alternative solution methods, and then analyze the critical weather scenario screening results.

\subsection{Experimental Setup}
\label{sec:exp_setup}
The adopted dataset consists of $T = 8784$ hourly weather--voltage pairs for the year 2016. The weather inputs are constructed from ERA5 reanalysis data~\cite{hersbach2023era5} at 701 grid points within the Texas grid boundary, covering three meteorological features: temperature, wind speed, and GHI. These weather conditions are directly incorporated into PowerWorld power flow simulations on the 6717-bus synthetic Texas system~\cite{overbye2023approach, safdarian2024calculation} to generate the corresponding voltages in per unit (p.u.). 
The resulting input dimension is $W = 2103$, and all inputs and outputs are normalized to $[0,1]$ using min-max scaling.

For the W2V model, the voltage-weighted loss in Section~\ref{sec:w2v_model} uses $\alpha = 10.3$ and $p = 2$, selected based on a trade-off analysis between overall and HV-specific voltage prediction accuracy. The latent space has dimension $K = 31$, initialized by 2 shared PCs between temperature and GHI, 4 temperature, 4 GHI, and 21 wind PCs. Once trained, the W2V model parameters are fixed throughout the screening process. Additional hyperparameters for the optimization and screening are summarized in Table~\ref{tab:params}.


\begin{table}[t]
\renewcommand{\arraystretch}{1.05}
\centering
\caption{Hyperparameters for the screening framework}
\vspace{-2mm}
\label{tab:params}
\begin{tabular}{lc}
\hline
\textbf{Parameter} & \textbf{Value} \\
\hline
HV voltage threshold $\tau$ & 1.20 p.u. \\
Top-$k$ buses for $S$ & 67 (top 1\%) \\
Classification threshold $\gamma$ & 1.186 \\
Perturbation budget $\epsilon$ & 0.3 \\
Primal step size $\eta_{\mathrm{p}}$ & $5 \times 10^{-2}$ \\
Dual step size $\eta_{\mathrm{d}}$ & $5 \times 10^{-1}$ \\
Max iterations & 5000 \\
\hline
\end{tabular}
\vspace{-3mm}
\end{table}

\subsection{Solution Method Comparison}
\label{subsec:method_comparison}
We compare the proposed solution method against two alternative configurations to validate the algorithmic design choices. The first alternative is a hierarchical primal--dual scheme that first optimizes the primal variable to convergence and then updates the dual variable, which isolates the benefit of simultaneous primal--dual updates. The second one replaces the L1 budget $\|\mathbf{u}\|_1 \leq \epsilon$ with an L2 constraint $\|\mathbf{u}\|_2 \leq \epsilon$, which isolates the effect of sparsity on solution quality and interpretability. Table~\ref{tab:method_comparison} summarizes the averaged results over 20 candidate scenarios.

\begin{table}[t]
\centering
\caption{Solution method comparison}
\vspace{-2mm}
\label{tab:method_comparison}
\begin{tabular}{lcc}
\hline
& Proposed & Hierarchical \\
\hline
$S^\star$ [p.u.] & 1.193 & 1.187 \\
Computation time [s] & 24.0 & 177.9 \\
\hline
& L1 (proposed) & L2 \\
\hline
$S^\star$ [p.u.] & 1.193 & 1.199 \\
$n_{\text{active}}$ & 4.2 & 30.9 \\
$|\bar{\Delta}T|$ & 0.025 & 0.027 \\
$|\bar{\Delta}W|$ & 0.083 & 0.065 \\
$|\bar{\Delta}G|$ & 0.020 & 0.023 \\
\hline
\end{tabular}

\vspace{2pt}
{\footnotesize All values are averaged over 20 candidate scenarios.}
\vspace{-4mm}
\end{table}

The proposed simultaneous scheme achieves higher $S^\star$ on average while reducing computation time by $7.4\times$ compared to the hierarchical approach. This speedup arises because the simultaneous updates allow the dual variables to track primal progress continuously, avoiding repeated primal iterations under outdated dual estimates. Since the screening framework needs to evaluate a large number of candidate scenarios, this computational advantage is important for scalability.

For the L1 versus L2 comparison, the L2 formulation yields a slightly higher mean $S^\star$, suggesting that distributing perturbations can further increase the optimized voltage stress. However, this increase is obtained with substantially reduced sparsity, as the L2 formulation activates nearly all latent dimensions on average, as reflected by the number of active latent variables $n_{\text{active}}=30.9$. In contrast, the L1 formulation activates only 4.2 variables on average, yielding a much more concentrated perturbation. This distinction is important since the screening framework aims not only to rank scenarios by voltage stress, but also to identify the dominant weather directions associated with HV vulnerability. The mean absolute standardized perturbation in the recovered weather space for each weather feature, i.e. $|\bar{\Delta}T|$, $|\bar{\Delta}W|$, and $|\bar{\Delta}G|$ for temperature, wind speed, and GHI, further supports this interpretation. Under the L1 constraint, the optimized perturbation is clearly wind-dominated: $|\bar{\Delta}W|$ is $3.3\times$ and $4.2\times$ larger than $|\bar{\Delta}T|$ and $|\bar{\Delta}G|$, respectively. In contrast, the L2 formulation spreads the perturbation more broadly across latent coordinates and weather features, making the dominant physical driver less distinguishable. Therefore, although the L2 formulation can marginally improve $S^\star$, the L1 formulation is adopted because it provides a more appropriate balance between voltage-stress amplification and diagnostic interpretability for critical weather scenario screening.

\subsection{Critical Weather Scenario Screening Results}
\label{subsec:screening_results}

\begin{table*}[t]
\renewcommand{\arraystretch}{1.05}
\centering
\caption{Top 10 Candidate Scenarios Ranked by $S^\star$}
\vspace{-2mm}
\label{tab:results}
\begin{tabular}{cccccccc}
\hline
$t$ & $S(\mathbf{z})$ & $S^\star$ & $|\bar{\Delta}T|$ & $|\bar{\Delta}W|$ & $|\bar{\Delta}G|$ & $n_{\text{active}}~\text{(S, T, W, G)}$ & Voltage Stressed Region \\
\hline
4962 & 1.172 & 1.232 & 0.033 & 0.090 & 0.026 & 3 (1, 0, 2, 0)  & Panhandle (upper)$^\ddagger$ \\
6497 & 1.179 & 1.222 & 0.018 & 0.089 & 0.015 & 2 (0, 0, 2, 0) & Panhandle (upper)$^\ddagger$ \\
3758 & 1.185 & 1.211 & 0.032 & 0.141 & 0.035 & 2 (1, 0, 1, 0)& Panhandle (upper)$^\ddagger$ \\
2087 & 1.176 & 1.202 & 0.028 & 0.089 & 0.011 & 2 (1, 0, 1, 0) & Panhandle (upper)$^\ddagger$ \\
6257 & 1.180 & 1.201 & 0.017 & 0.060 & 0.010 & 5 (1, 0, 4, 0) & Panhandle (upper)$^\ddagger$ \\
90   & 1.171 & 1.197 & 0.032 & 0.184 & 0.029 & 1 (0, 0, 1, 0) & Panhandle (lower)$^\ddagger$ \\
3784 & 1.167 & 1.194 & 0.023 & 0.124 & 0.021 & 3 (1, 0, 2, 0)& Panhandle (lower)$^\ddagger$ \\
7931 & 1.178 & 1.193 & 0.005 & 0.030 & 0.004 & 13 (2, 1, 9, 1)& West Texas$^\ddagger$ \\
8486 & 1.177 & 1.190 & 0.010 & 0.052 & 0.007 & 7 (0, 1, 5, 1)& Panhandle (lower) \\
5484 & 1.173 & 1.187 & 0.004 & 0.019 & 0.004 & 12 (2, 1, 8, 1)& Panhandle (lower) \\
\hline
\end{tabular}
\vspace{2pt}

{\footnotesize S/T/W/G: number of active latent variables initialized by shared, temperature, wind, and GHI PCs, respectively.}
\vspace{-2mm}
\end{table*}

\begin{figure*}[t]
\centering
\setlength{\tabcolsep}{1pt}
\begin{tabular}{@{}ccc@{}}
\includegraphics[width=0.325\textwidth]{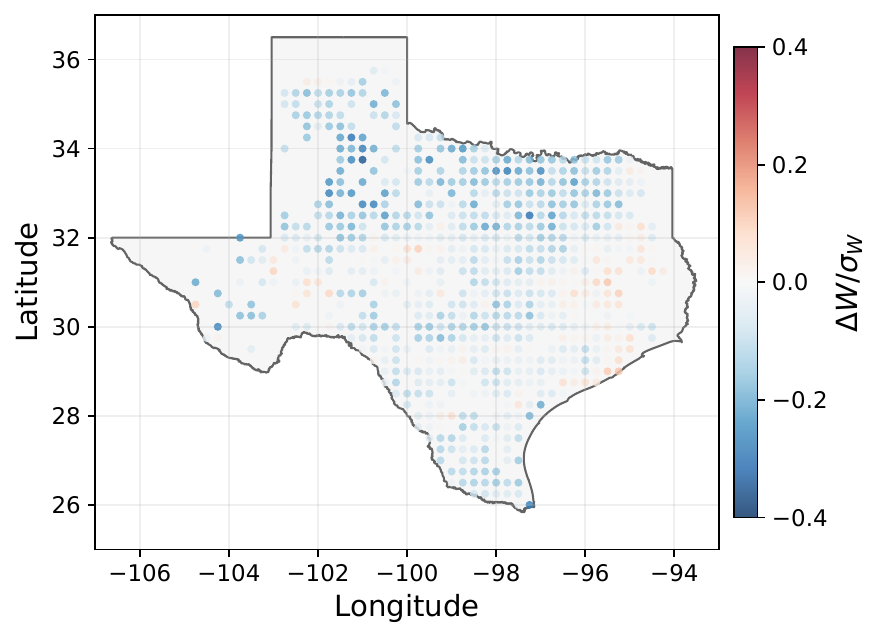} &
\includegraphics[width=0.325\textwidth]{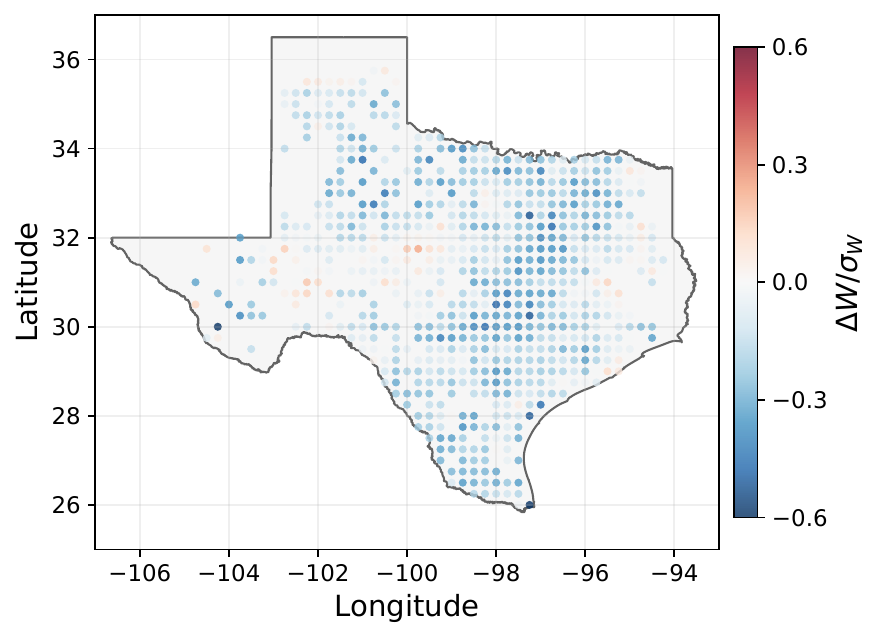} &
\includegraphics[width=0.325\textwidth]{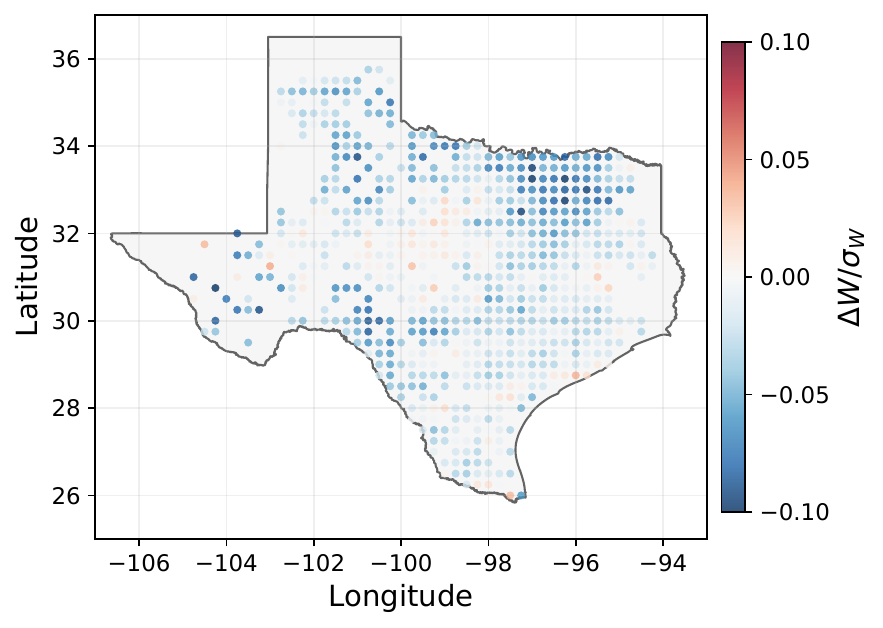} \\[-2pt]
\includegraphics[width=0.325\textwidth]{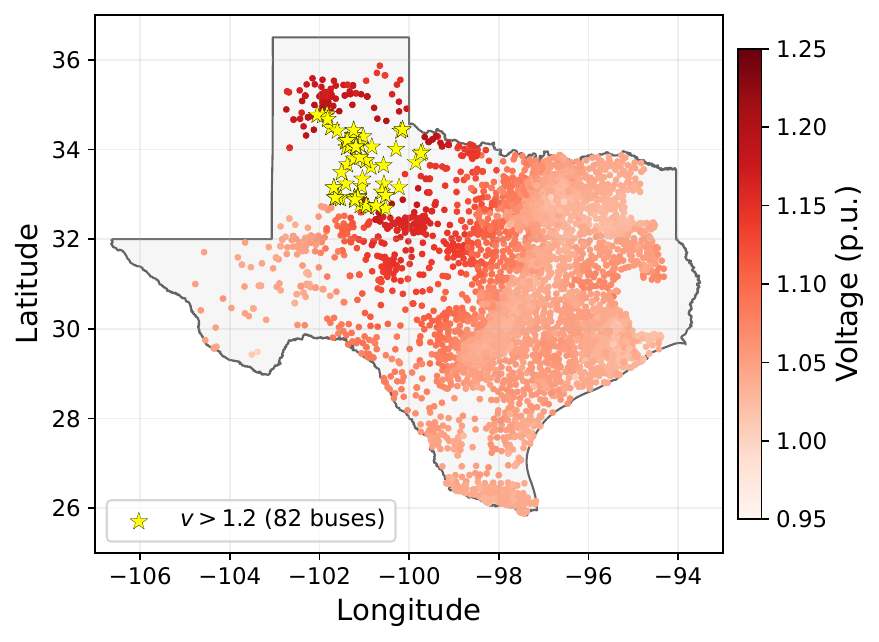} &
\includegraphics[width=0.325\textwidth]{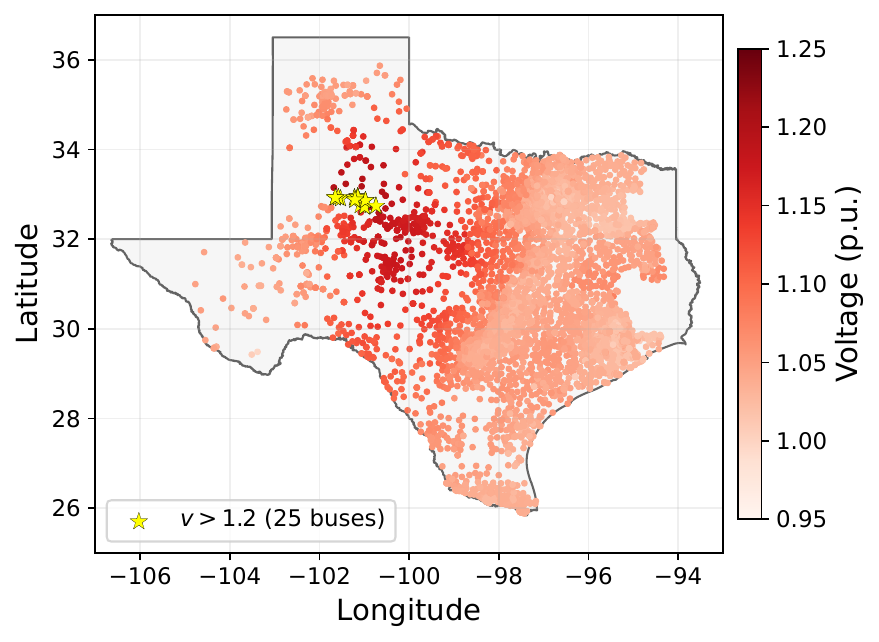} &
\includegraphics[width=0.325\textwidth]{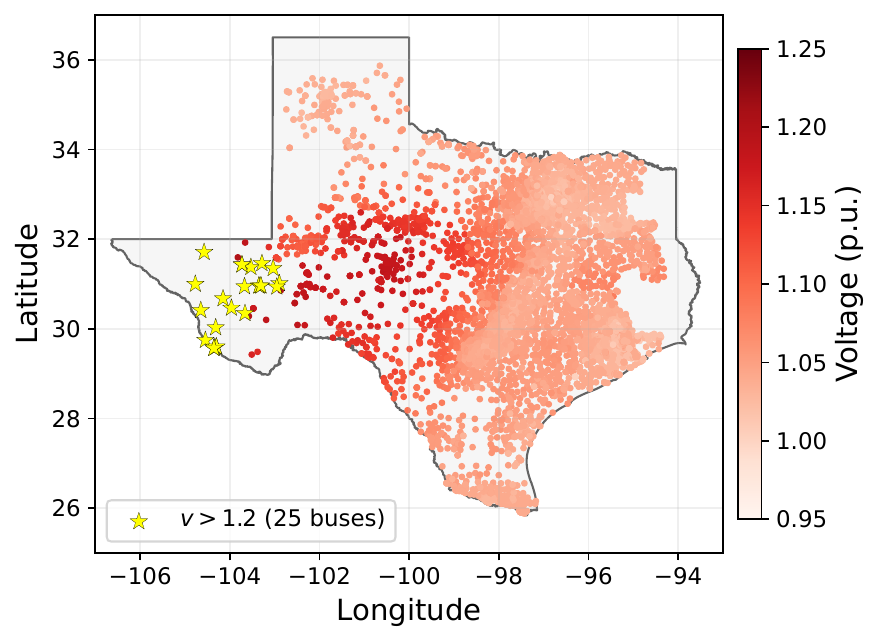} \\[1pt]
{\small (a) $t=6497$ (Upper Panhandle)} &
{\small (b) $t=90$ (Lower Panhandle)} &
{\small (c) $t=7931$ (West Texas)} \\
\end{tabular}
\caption{Standardized wind speed perturbation for 701 grid points (top row) and corresponding voltage distribution across 6717 buses (bottom row) for three representative scenarios from distinct stressed regions.}
\label{fig:perturbation_voltage}
\vspace{-4mm}
\end{figure*}

Table~\ref{tab:results} presents the top 10 candidate scenarios ranked by the optimal criticality score $S^\star$. For each scenario, the table reports the initial criticality score $S(\mathbf{z})$, the mean absolute standardized perturbation per weather feature ($|\bar{\Delta}T|$, $|\bar{\Delta}W|$, $|\bar{\Delta}G|$), the number of active latent variables with their composition by PCA initialization, and the geographic region exhibiting the highest voltage stress. 
The $^\ddagger$ marker indicates scenarios where bus voltages exceeding $\tau$ are observed under the optimized perturbation.

A key observation in Table~\ref{tab:results} is that the final ranking by $S^\star$ does not simply follow the initial criticality score $S(\mathbf{z})$. For example, the scenario at $t=4962$ achieves the highest optimized score although its initial score is lower than that of $t=3758$. This indicates that vulnerability is not determined solely by the initial distance to the HV boundary. Instead, it also depends on the local sensitivity of weather-to-voltage mapping around each scenario. This sensitivity-aware ranking is precisely the purpose of the proposed optimization-based screening framework.

The results also show that wind speed perturbations are the largest among weather features across all scenarios. 
This indicates that wind speed is the most influential factor for triggering HV stress in the considered Texas system. The sparsity pattern of the active latent variables also supports the same conclusion. Most scenarios activate only a small number of latent variables ($n_{\text{active}}$), typically between 2 and 5, and the active set always includes at least one wind-initialized latent component. Therefore, HV vulnerability is driven by a small number of dominant wind-related latent directions, rather than perturbations across all weather features.

The geographic distribution of the voltage-stressed regions provides additional physical insight. The top 5 scenarios are concentrated in the upper Panhandle region, where wind generation buses are most densely located in the synthetic Texas system. Lower-ranked scenarios shift toward the lower Panhandle and West Texas. This pattern suggests that wind-dominated perturbations translate into voltage stress most effectively in regions with high wind generation concentration.

Fig.~\ref{fig:perturbation_voltage} illustrates the spatial distribution of the standardized wind speed perturbation and the resulting voltage response from three distinct stressed regions. While wind speed perturbations dominate in all cases, their spatial pattern varies by scenario. In each case, a clear wind speed reduction is observed in the corresponding voltage-stressed region, and the optimization preferentially reduces wind speed at locations where the original wind condition is already low. When the original wind speed is already close to zero, optimization instead reduces the wind speed in surrounding locations, collectively driving voltage stress in the nearby region. This behavior indicates that the framework can produce a spatially coordinated perturbation, adaptive to the local vulnerability structure of each scenario.

\begin{figure}[t]
\centering
\setlength{\tabcolsep}{1pt}
\begin{tabular}{@{}cc@{}}
\includegraphics[width=0.49\columnwidth]{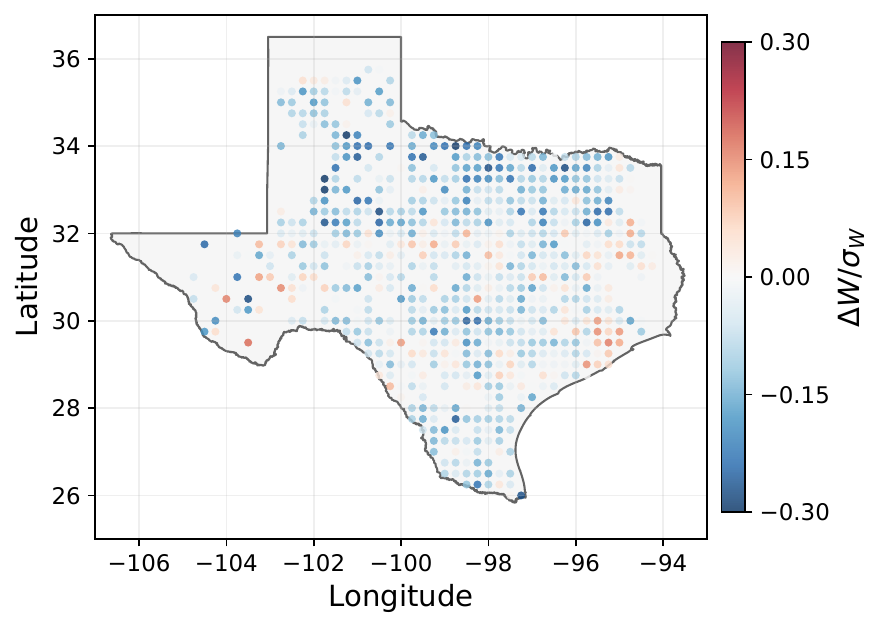} &
\includegraphics[width=0.49\columnwidth]{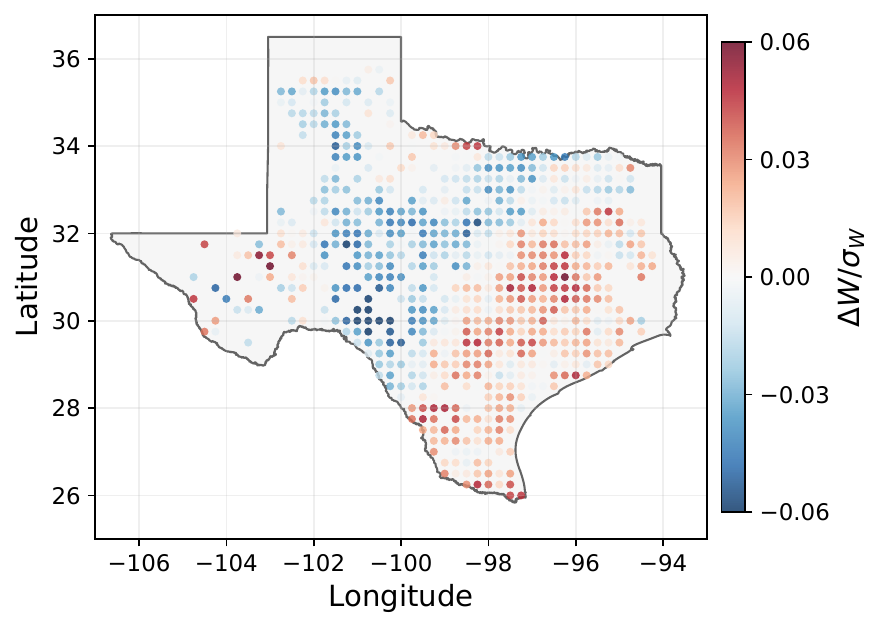} \\[-2pt]
\includegraphics[width=0.49\columnwidth]{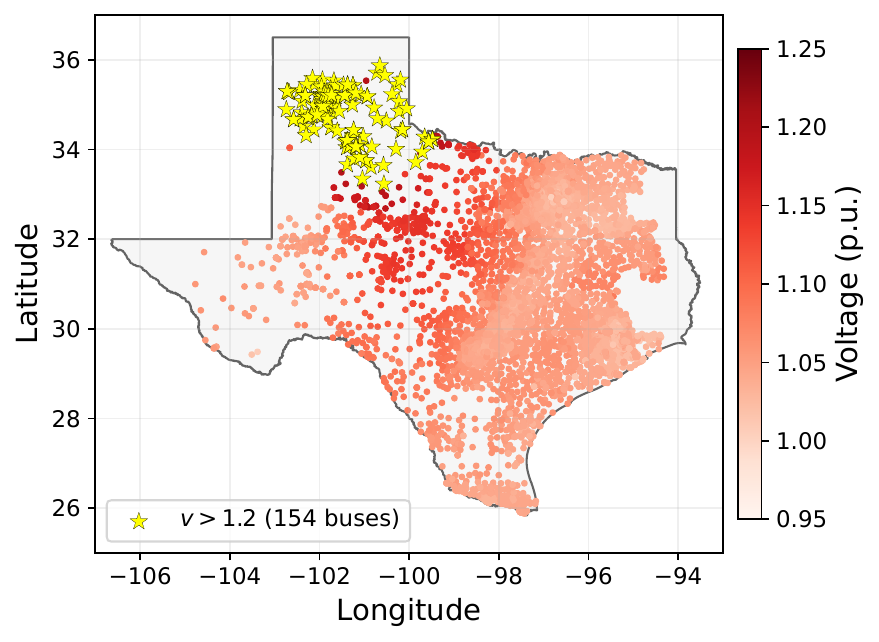} &
\includegraphics[width=0.49\columnwidth]{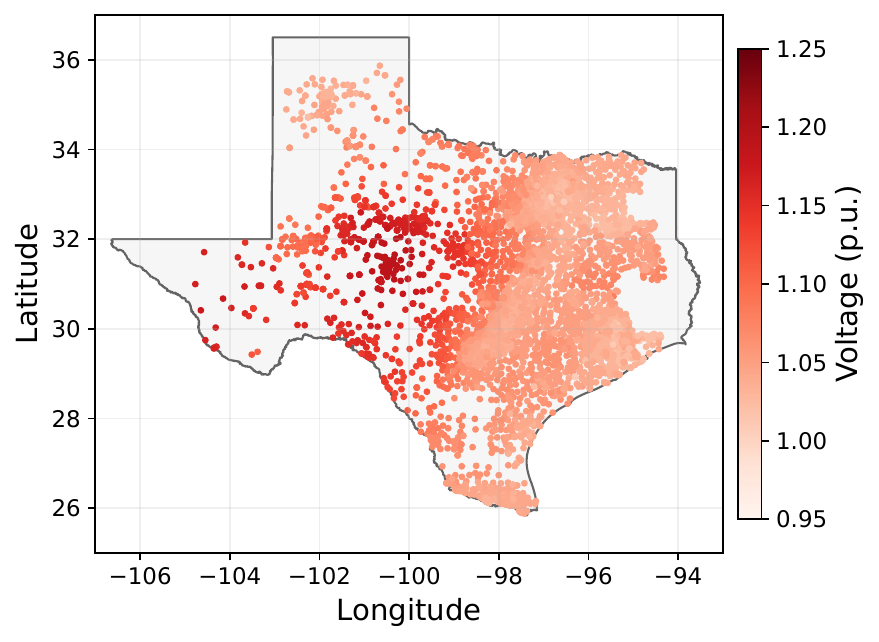} \\[1pt]
{\small (a) $t=4962$} &
{\small (b) $t=5484$} \\
\end{tabular}
\vspace{-3mm}
\caption{Comparison of two scenarios with similar $S(\mathbf{z})$ but different $S^\star$.}
\label{fig:case_pair}
\vspace{-5mm}
\end{figure}

A focused pairwise example in Fig.~\ref{fig:case_pair} further illustrates why the proposed method can provide information beyond the initial criticality score. The scenarios at $t=4962$ and $t=5484$ have nearly identical initial criticality scores, $S(\mathbf{z})=1.172$ and $1.173$, respectively. However, their optimized scores diverge substantially, with $S^\star=1.232$ and $1.187$. In the former case, the optimized perturbation concentrates wind speed reduction in the upper Panhandle with a compact active set of three latent variables, leading to bus voltages exceeding $\tau$ at multiple buses. In contrast, the latter requires a broader active set of twelve latent variables, but the resulting voltage stress centered in the lower Panhandle remains lower and does not exceed $\tau$. This comparison shows that nearly identical initial scores can result in significantly different vulnerability under the same perturbation budget.

To assess whether the identified perturbation patterns are physically meaningful, we examine the actual HV events in the dataset. Specifically, we isolate timesteps satisfying $\max_i v_{t,i} > \tau$ and analyze their corresponding weather conditions and voltage-stressed regions. These HV timesteps are generally associated with low wind-speed conditions, often accompanied by elevated temperature and GHI. Geographically, HV buses appear consistently in the Panhandle and West Texas regions, coinciding with areas of high wind generation capacity. This empirical pattern aligns well with the perturbation direction identified by the proposed framework, where wind speed reductions are the dominant mechanism for moving non-critical scenarios toward HV stress. Since these HV events are identified independently of the W2V model, this consistency provides corroborating evidence that the proposed screening results capture physically meaningful weather-voltage sensitivities rather than artifacts of the model.

\section{Conclusion}
\label{sec:conclusion}
This paper has developed a critical weather scenario screening framework that leverages the differentiable W2V predictive model to identify weather conditions that could trigger power grid HV events. By optimizing the latent-space perturbation under a uniform budget across candidate scenarios, our approach provides a vulnerability metric for possible weather scenarios that incorporates the weather-to-voltage sensitivity in determining the proximity to HV events.
Numerical experiments on the 6717-bus synthetic Texas grid have identified HV-vulnerable scenarios that are not fully captured by the initial voltage analysis alone. The top-ranked scenarios exhibit sparse, wind-dominated perturbation patterns, with voltage stress concentrated primarily in regions with high penetration of wind generation. These findings are consistent with the occurrence of actual HV events that have been observed in the dataset. Together, these results demonstrate the capability in efficient, diagnostic screening for weather-induced voltage risk assessment. This can, in turn, help grid operators develop screening-informed operating rules to quickly forecast and proactively prepare for stressed grid states. As sources of grid stress and voltage risk continue to diversify, we plan to expand our current weather-induced framework to generalize more external disturbances such as cyberattacks and human factors.


\bibliographystyle{IEEEtran}
\bibliography{mybib}

\end{document}